\documentclass[preprint,12pt]{elsarticle}

\usepackage{ifthen} 
\newboolean{pdflatex}
\setboolean{pdflatex}{true} 

\newboolean{articletitles}
\setboolean{articletitles}{true} 

\newboolean{uprightparticles}
\setboolean{uprightparticles}{false} 

\newboolean{inbibliography}
\setboolean{inbibliography}{false} 

\usepackage{booktabs}
\usepackage[utf8]{inputenc}
\usepackage{xspace}
\usepackage[separate-uncertainty = true]{siunitx}
\usepackage{hhline}
\usepackage{collcell}
\usepackage{tabularx}
\usepackage{amssymb}
\usepackage{amsmath}
\usepackage{amsfonts}
\usepackage{comment}
\usepackage{lineno}
\usepackage[colorinlistoftodos]{todonotes}
\usepackage{makecell}
\usepackage{microtype}
\usepackage{caption} 
\usepackage{graphicx}
\usepackage{color}
\usepackage{colortbl}
\usepackage{pdflscape}
\usepackage{upgreek} 
\usepackage{natbib}
\usepackage{cleveref}
\usepackage{subcaption}

\graphicspath{{./figs/}} 
\usepackage[nobiblatex]{xurl}

\usepackage{xspace} 
\usepackage{upgreek}

\def\MagUp {\mbox{\em Mag\kern -0.05em Up}\xspace}

\ifthenelse{\boolean{uprightparticles}}%
{

 \def\PDelta      {\ensuremath{\Delta}\xspace}                 
 \def\PXi         {\ensuremath{\Xi}\xspace}                 
 \def\PLambda     {\ensuremath{\Lambda}\xspace}                 
 \def\PSigma      {\ensuremath{\Sigma}\xspace}                 
 \def\POmega      {\ensuremath{\Omega}\xspace}                 
 \def\PUpsilon    {\ensuremath{\Upsilon}\xspace}
 \let\oldPi\Pi
 \def\PPi         {\ensuremath{\oldPi}\xspace}

 \def\PB      {\ensuremath{\mathrm{B}}\xspace}                 
                  
 \def\PD      {\ensuremath{\mathrm{D}}\xspace}

 \def\PK      {\ensuremath{\mathrm{K}}\xspace}

 \def\Pe      {\ensuremath{\mathrm{e}}\xspace}

 \def\Pi      {\ensuremath{\mathrm{i}}\xspace}

 \def\Ps      {\ensuremath{\mathrm{s}}\xspace}

 \def\thebaroffset{0.0em}
}
{

 \mathchardef\PDelta="7101
 \mathchardef\PXi="7104
 \mathchardef\PLambda="7103
 \mathchardef\PSigma="7106
 \mathchardef\POmega="710A
 \mathchardef\PUpsilon="7107
 \mathchardef\PPi="7105
                  
 \def\PB      {\ensuremath{B}\xspace}                 
                  
 \def\PD      {\ensuremath{D}\xspace}

 \def\PK      {\ensuremath{K}\xspace}

 \def\Pe      {\ensuremath{e}\xspace}

 \def\Pi      {\ensuremath{i}\xspace}

 \def\Ps      {\ensuremath{s}\xspace}

 \def\thebaroffset{0.18em}
}
\newcommand{\offsetoverline}[2][\thebaroffset]{\kern #1\overline{\kern -#1 #2}}%

\makeatletter
\ifcase \@ptsize \relax
  \newcommand{\miniscule}{\@setfontsize\miniscule{4}{5}}
\or
  \newcommand{\miniscule}{\@setfontsize\miniscule{5}{6}}
\or
  \newcommand{\miniscule}{\@setfontsize\miniscule{5}{6}}
\fi
\makeatother

\DeclareRobustCommand{\optbar}[1]{\shortstack{{\miniscule (\rule[.5ex]{1.25em}{.18mm})}
  \\ [-.7ex] $#1$}}

\def\en         {{\ensuremath{\Pe^-}}\xspace}   

\def\squark    {{\ensuremath{\Ps}}\xspace}

\def\KorKbar {\kern \thebaroffset\optbar{\kern -\thebaroffset \PK}{}\xspace}

\def\D       {{\ensuremath{\PD}}\xspace}

\def\DorDbar {\kern \thebaroffset\optbar{\kern -\thebaroffset \PD}\xspace}

\def\Dp      {{\ensuremath{\D^+}}\xspace}
\def\Dm      {{\ensuremath{\D^-}}\xspace}

\def\DpDm    {\ensuremath{\Dp {\kern -0.16em \Dm}}\xspace}

\def\B       {{\ensuremath{\PB}}\xspace}

\def\BorBbar {\kern \thebaroffset\optbar{\kern -\thebaroffset \PB}\xspace}

\def\Bd      {{\ensuremath{\B^0}}\xspace}

\def\BdorBdbar {\kern \thebaroffset\optbar{\kern -\thebaroffset \Bd}\xspace}

\def\Bs      {{\ensuremath{\B^0_\squark}}\xspace}

\def\BsorBsbar {\kern \thebaroffset\optbar{\kern -\thebaroffset \Bs}\xspace}

\def\Y#1S{\ensuremath{\PUpsilon{(#1S)}}\xspace}

\def\LorLbar     {\kern \thebaroffset\optbar{\kern -\thebaroffset \PLambda}\xspace}

\def\AT#1     {\ensuremath{A_{\mathrm{T}}^{#1}}\xspace}           

\def\C#1      {\ensuremath{\mathcal{C}_{#1}}\xspace}                       
\def\Cp#1     {\ensuremath{\mathcal{C}_{#1}^{'}}\xspace}                    
\def\Ceff#1   {\ensuremath{\mathcal{C}_{#1}^{\mathrm{(eff)}}}\xspace}        
\def\Cpeff#1  {\ensuremath{\mathcal{C}_{#1}^{'\mathrm{(eff)}}}\xspace}       
\def\Ope#1    {\ensuremath{\mathcal{O}_{#1}}\xspace}                       
\def\Opep#1   {\ensuremath{\mathcal{O}_{#1}^{'}}\xspace}                    

\newcommand{\nospaceunit}[1]{\ensuremath{\text{#1}}}       
\newcommand{\aunit}[1]{\ensuremath{\text{\,#1}}}       

\newcommand{\tev}{\aunit{Te\kern -0.1em V}\xspace}
\newcommand{\gev}{\aunit{Ge\kern -0.1em V}\xspace}
\newcommand{\mev}{\aunit{Me\kern -0.1em V}\xspace}
\newcommand{\kev}{\aunit{ke\kern -0.1em V}\xspace}
\newcommand{\ev}{\aunit{e\kern -0.1em V}\xspace}
 
\newcommand{\mevc}{\ensuremath{\aunit{Me\kern -0.1em V\!/}c}\xspace}
\newcommand{\gevc}{\ensuremath{\aunit{Ge\kern -0.1em V\!/}c}\xspace}
\newcommand{\mevcc}{\ensuremath{\aunit{Me\kern -0.1em V\!/}c^2}\xspace}
\newcommand{\gevcc}{\ensuremath{\aunit{Ge\kern -0.1em V\!/}c^2}\xspace}

\def\cm   {\aunit{cm}\xspace}

\def\mm   {\aunit{mm}\xspace}

\def\mum  {\ensuremath{\upmu\nospaceunit{m}}\xspace}

\def\ns   {\ensuremath{\aunit{ns}}\xspace}
\def\ps   {\ensuremath{\aunit{ps}}\xspace}

\def\Xrad {\ensuremath{X_0}\xspace}

\def\gsim{{~\raise.15em\hbox{$>$}\kern-.85em
          \lower.35em\hbox{$\sim$}~}\xspace}
\def\lsim{{~\raise.15em\hbox{$<$}\kern-.85em
          \lower.35em\hbox{$\sim$}~}\xspace}

\def\degrees{\ensuremath{^{\circ}}\xspace}

\def\mrad{\aunit{mrad}\xspace}

\def\nonp {\ensuremath{\mathrm{{ \mathit{n^+}} \mbox{-} on\mbox{-}{ \mathit{p}}}}\xspace}

\def\tell1  {TELL1\xspace}
\def\ukl1   {UKL1\xspace}

\def\ke {\ensuremath{ \text{k}e^{-} }\xspace}

\def\XResThick{\ensuremath{3.3  \pm 0.3~\mum}\xspace}
\def\XResThin {\ensuremath{14.4 \pm 0.5~\mum}\xspace}
\def\YResThick{\ensuremath{3.5  \pm 0.3~\mum}\xspace}
\def\YResThin {\ensuremath{14.3 \pm 0.5~\mum}\xspace}
\def\XResTrack{\ensuremath{2.3  \pm 0.1~\mum}\xspace}
\def\YResTrack{\ensuremath{2.4  \pm 0.1~\mum}\xspace}
\def\TrackTimeRes{\ensuremath{92 \pm 5~\ps}\xspace}
\def\ThickTimeRes{\ensuremath{554 \pm 16~\ps}\xspace}
\def\TRefRes{\ensuremath{12~\ps}\xspace}
\def\ThinTimeRes{\ensuremath{184\pm5~\ps}\xspace}
\def\SingleTimeRes{\ensuremath{175 \pm 5~\ps}\xspace}

\def\ps     {\ensuremath{\aunit{ps}}\xspace}
\def\kepler {\textsc{Kepler}\xspace}
\newcommand {\mv}{\aunit{m\kern -0.1em V}\xspace}
\newcommand {\kv}{\aunit{k\kern -0.1em V}\xspace}

\graphicspath{{./figs/}}

\begin{document}
\begin{frontmatter}
\title{The Timepix4 Beam Telescope}

\author[a]{K.~Akiba\corref{cor1}}
\author[b]{J.~Alozy}
\author[c]{D.~Bacher}
\author[b]{R.~Ballabriga~Sune}
\author[b]{F.~de~Benedetti}
\author[a]{M.~van~Beuzekom}
\author[a]{V.~van~Beveren}
\author[a]{T.~Bischoff}
\author[a]{H.~Boterenbrood}
\author[b]{W.~Byczynski}
\author[b]{M.~Campbell}
\author[a]{E.~Chatzianagnostou}
\author[b]{V.~Coco}
\author[b]{P.~Collins}
\author[b,d]{E.~Dall'Occo}
\author[b]{R.~Dumps}
\author[b]{F.~de~Benedetti}
\author[a,e]{T.~Evans}
\author[f]{A.~Fernández~Prieto}
\author[a]{M.~Fransen}
\author[f]{A.~Gallas~Torreira}
\author[c]{R.~Gao}
\author[a]{R.~Geertsema}
\author[c]{F.~Gonçalves~Abrantes}
\author[a]{V.~Gromov}
\author[b]{M.~M.~Halvorsen}
\author[a]{B.~van~der~Heijden}
\author[a]{K.~Heijhoff}
\author[c]{M.~John}
\author[g]{D.~Johnson}
\author[a]{U.~Krämer}
\author[b]{E.~Lemos~Cid}
\author[b]{X.~Llopart~Cudie}
\author[g]{M.~J.~Madurai}
\author[a]{D.~Oppenhuis}
\author[b]{T.~Pajero}
\author[f]{E.~Rodríguez~Rodríguez}
\author[d]{D.~Rolf}
\author[a]{A.~Sarnatskiy}
\author[b]{H.~Schindler}
\author[f]{P.~Vázquez~Regueiro} 
\author[a]{A.~Vitkovksiy}
\author[b]{M.~Williams}
\author[a,h]{G.~Wang}

\cortext[cor1]{Corresponding author}

\affiliation[a]{{Nikhef, Science Park 105, 1098 XG Amsterdam, the Netherlands}}
\affiliation[b]{{CERN, Esplanade des Particules 1, 1211 Geneva, Switzerland}}
\affiliation[c]{{Department of Physics, University of Oxford, Denys Wilkinson Bldg., Keble Road, Oxford, OX1~3RH, United Kingdom}}
\affiliation[d]{{TU Dortmund, Otto-Hahn-Strasse 4, 44227 Dortmund, Germany}}
\affiliation[e]{{Department of Physics and Astronomy, University of Manchester, Manchester, United Kingdom}}
\affiliation[f]{{Instituto Galego de Fisica de Altas Enerxias (IGFAE), Universidade de Santiago de Compostela, Santiago de Compostela, Spain}}
\affiliation[g]{{School of Physics and Astronomy, University of Birmingham, Edgbaston, Birmingham, B15 2TT, United Kingdom}}
\affiliation[h]{{Central China Normal University, Luoyu Road 152, Hongshan District, Wuhan, China}}


\begin{abstract}
The spatial and temporal performance of a telescope system comprising planar silicon sensors bump-bonded to Timepix4-v2 ASICs are assessed at the CERN SPS using a $180$~GeV/$c$ mixed hadron beam.
The pointing resolution at the centre of the telescope is 
\XResTrack and \YResTrack in  x and y directions, respectively.  
The temporal resolution for the combined timestamps on a track using only the information from the Timepix4 planes is found to be \TrackTimeRes.
The telescope is extensively described including its data acquisition system and dedicated software as well as the corrections applied to the raw measurements.
\end{abstract}


\end{frontmatter}




\section{Introduction} 
\label{sec:introduction} 
Precise timing measurements can be added to high-granularity pixel detectors in order to improve the reconstruction of charged-particle trajectories, especially in high-multiplicity environments.
This approach, also known as 4D-tracking, is being developed in high-energy-physics experiments as a solution to operate the detectors at high instantaneous luminosities.
These measurements are crucial for distinguishing vertices produced in nearly simultaneous collisions, as described in~\cite{LHCbVELOgroup:2022}.
A telescope system has been implemented using pixelated silicon sensors bonded to version 2 of the Timepix4 Application Specific Integrated Circuit (ASIC). 
The main objective is to demonstrate high-rate 4D tracking and achieve combined track timestamps with superior temporal resolution compared to single-hit measurements.
The precise four-dimensional pointing resolution can be used for detailed characterisation of advanced detectors with high spatial and temporal precision. 

The Timepix ASIC family has previously been employed in reconstructing the trajectories of charged particles in several experiments and as an R\&D platform for sensors, ASICs, and other detector components essential for the LHCb experiment upgrade~\cite{Akiba:2019faz,Heijhoff:2020mlk,Heijhoff:2021rtu,DallOcco:2021tjb}. 
The Timepix4~\cite{tpx4_jinst} is designed to reach an excellent temporal resolution of around $60\ps$~\cite{Heijhoff_2022}, in addition to spatial measurements with micrometre precision. 

This paper outlines the design of an eight-plane telescope based on the Timepix4 v2 ASIC. 
It also describes the data acquisition system, operational infrastructure, and dedicated software employed in the setup. 
The spatial and temporal resolutions of this system are measured using a mixed hadron beam with an energy of 180 GeV at the SPS H8 beam line facility~\cite{sps-h8}.

\section{Setup description} 
\label{sec:setup} 
\begin{figure}[tb]
	\centering
	\includegraphics[width=0.7\textwidth]{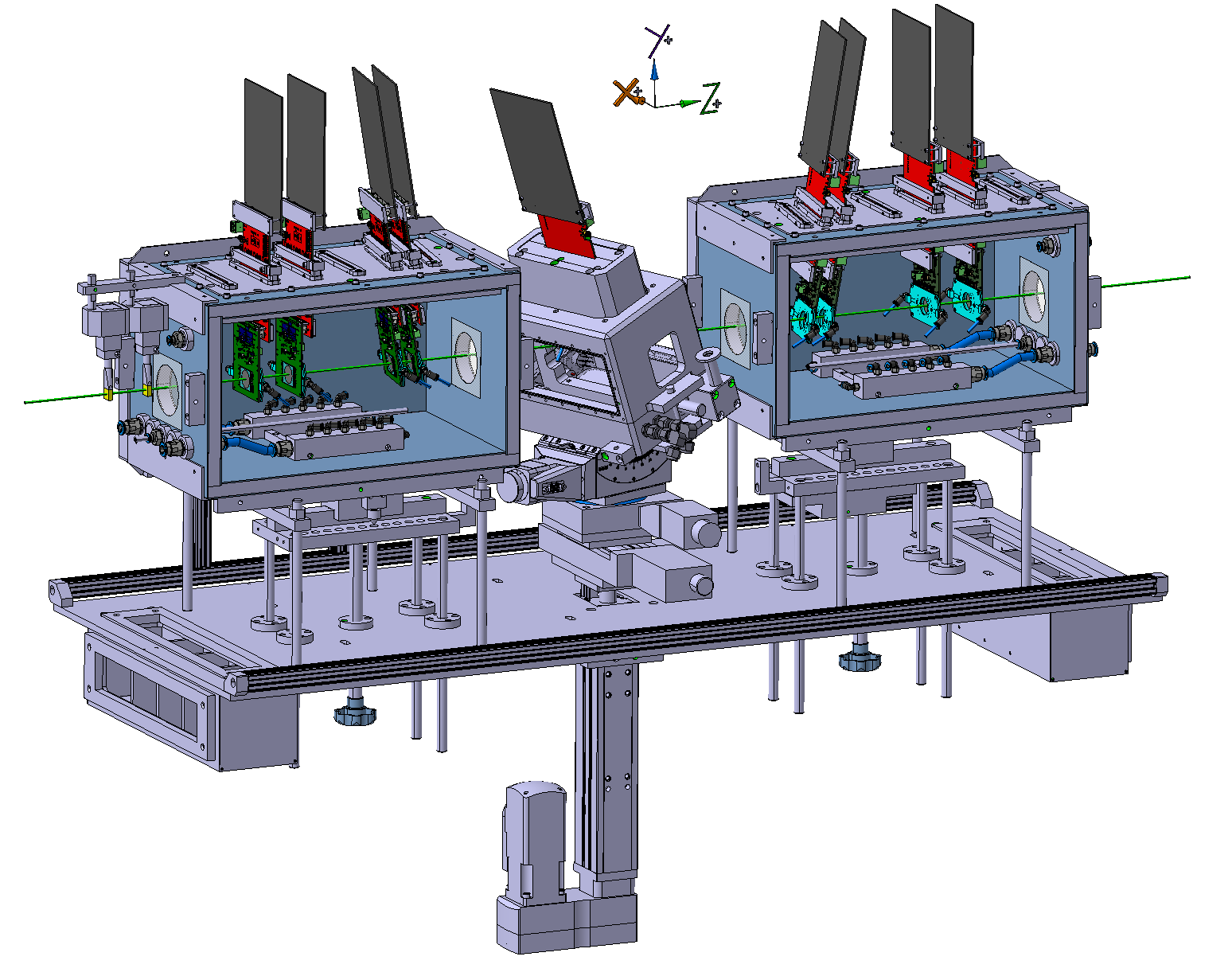}
	\caption{Mechanical design of the telescope arrangement of eight measuring planes and a device under test. 
    The solid line cutting through the planes represents the traversing beam, where particles travel from left to right.}
	\label{fig:telescope}
\end{figure}

\begin{figure}[tb]
	\centering
	\includegraphics[width=\textwidth]{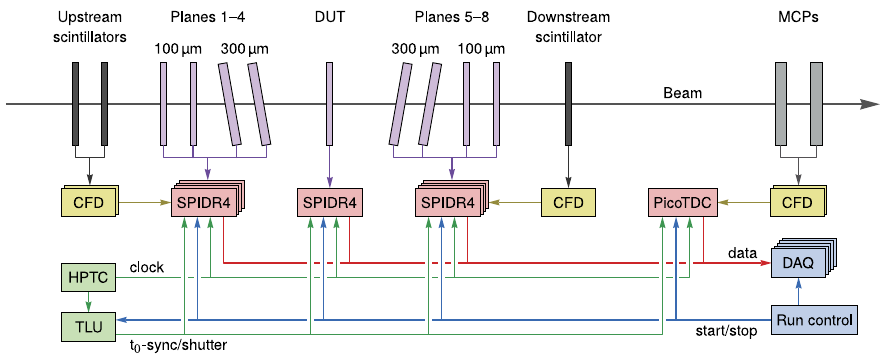}%
	\caption{Diagram of the Timepix4 Telescope.}
	\label{fig:telescopeDiagram}
\end{figure}

The telescope is composed of two arms, each comprising four detector planes, as depicted in \cref{fig:telescope}. 
A global right-handed coordinate system is defined where the $z$ axis is aligned with the direction of the beam, and the $y$ axis points upwards. 
A diagram of the telescope setup is shown in \cref{fig:telescopeDiagram}.

The detectors are composed of planar silicon pixel sensors bump bonded to Timepix4 ASICs.
They are mounted inside hermetic enclosures described in \cite{Akiba:2022fnt}. 
These boxes are supported by four rods to minimise vibrations.
The temperature and humidity inside the enclosures are continuously monitored.
The slots and corresponding flanges on the top covers are machined with high precision and define the angles of the planes with respect to the beam, as well as the relative distances between planes.
The outermost slots of each box have been designed for perpendicular planes and are populated with thin (100~$\mum$) sensors to achieve the best timing performance, while the remaining (300~$\mum$) sensors are angled at $9\degrees$ with respect to the $x$ and $y$ axes to improve charge sharing between adjacent pixels and obtain a better spatial resolution. 
With respect to the DUT, the positions of the telescope planes along the $z$ axis are at $-470$, $-420$, $-295$, $-265$, $325$, $355$, $480$ and $530\mm$. 
The base of the telescope is mounted on a remote-controlled motion stage in order to align the entire telescope with respect to the beam in the $x$ and $y$ directions.

\subsection{Timepix4 ASIC}
\label{sec:timepix4}
Timepix4 is the latest pixel readout ASIC for hybrid pixel detectors in the Medipix family~\cite{Ballabriga2020}. 
It was developed in a \SI{65}{\nano\meter} CMOS technology by CERN, Nikhef, and IFAE. The pixel matrix consists of $448\times512$~square pixels with a pitch of \SI{55}{\micro\meter}, and the front-end electronics of the pixels consists of a charge-sensitive preamplifier~\cite{Ballabriga2022} with the ability to absorb leakage current from the sensor and a per-pixel threshold adjustment followed by a leading-edge discriminator. 
The Equivalent Noise Charge (ENC) of the preamplifier depends on the ASIC settings and sensor capacitance and is typically about $80$~electrons. The device is usually operated with a threshold of about $1000$~electrons. 
The ASICs are configured to register both the Time of Arrival~(ToA) and the Time over Threshold~(ToT) of each discriminated signal. 
The ToT is a measure of the amount of charge in the signal delivered by the sensor (\cref{sec:chargecalib}) and it is used to improve both the temporal resolution (\cref{sec:timewalkCorrection}) and the spatial resolution. 

The ToA and ToT are measured by a time-to-digital converter~(TDC),  using a Voltage-Controlled Oscillator~(VCO) with a nominal frequency of \SI{640}{\mega\hertz}. 
The VCO is shared by a group of two by four pixels, referred to as superpixel. 
The ToA timestamp is given by the following expression: 
 \begin{equation}
    \textnormal{ToA} = \textnormal{cToA}\times\textnormal{T$_{\textnormal{clock}}$}  - \textnormal{fToA}\times\textnormal{T$_{\textnormal{VCO}}$} - \textnormal{ufToA}\times\frac{\textnormal{T$_{\textnormal{VCO}}$}}{8}.
    \label{eq:toa}
\end{equation}
Here cToA is a counter of the number of clock ticks of the 40 MHz\footnote{ The actual clock is 40.08~MHz, which is the clock frequency of the LHC.} system clock, with a corresponding period time (T$_{\textnormal{clock}}$) of \SI{25}{\nano\second}.
The second term, fToA (fine ToA) counts the number of rising edges of the VCO clock, from the arrival of the hit to the first subsequent rising edge of the system clock. 
The nominal period of the VCO (T$_{\textnormal{VCO}}$) is around  \SI{1.56}{\nano\second}, with the exact value varying between the superpixels. 
This is discussed in more detail in \cref{sec:vco_correction}. 
The ultra-fine ToA, ufToA, is a further refinement of the measurement using four phase-shifted copies of the VCO clock and has a granularity of T$_{\textnormal{VCO}}$/8.

\subsection{Sensors}
The sensors are manufactured by ADVAFAB and consist of high resistivity $p$-type silicon with segmented $n^+$ pixel implants and a non-segmented $p^+$ implant on the backside. 
The pixelated side has $448 \times 512$  square implants with a width of around $39~\mum$, separated by a uniform $p$-spray, and covered with under-bump metallisation which allows the pixel sensors to be bonded with solder bumps to the ASICs.
The 300~\mum sensors are fully depleted at a reverse bias voltage of approximately 50~V.
The 100~\mum thick sensors are depleted at around 10~{V}.
One of the four thin sensors breaks down at around 50~V while the others can sustain a voltage of at least 140~V.

\begin{figure}[tb]
	\centering
	\includegraphics[width=0.4\textwidth]{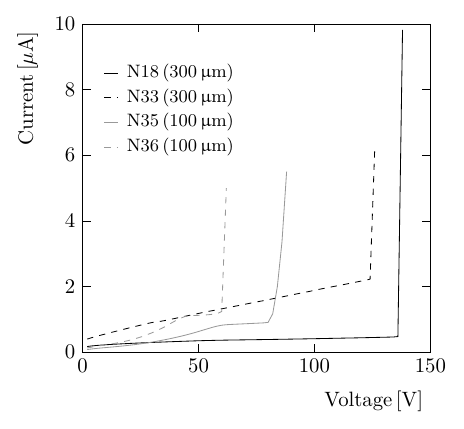} 
	\caption{ I-V characteristic curves for N18, N33, N35 and N36. }
	\label{fig:ivs}
\end{figure}

Laboratory characterisation measurements and TCAD simulation studies indicate that the observed sensor breakdown values are thought to occur due to a highly doped $p$-spray layer \cite{Haimberger:2023fkr}.
Four I-V characteristic curves of the 300~\mum and 100~\mum thick sensors are shown in \cref{fig:ivs}.

\subsection{Cooling}  
All sensor planes are cooled via direct contact with a cooling block attached to the backside of the PCB (chipboard) carrying the Timepix4. 
The cooling blocks are made of 3D-printed titanium with hollow cavities in which liquid glycol is circulated.
The cooling fluid is pumped through the cooling block by an off-the-shelf chiller at a temperature of \SI{18}{\celsius}, and is distributed in parallel to each of the planes. 
The cooling blocks have a circular cut-out to minimise the material traversed by the beam. 
The interface between the chipboard and its cooling block is improved by applying a high thermal conductivity sheet\footnote{Bergquist gap pad TGP 800VOS.}.

\subsection{Detector Planes}  
Each detector plane is composed of a sensor and ASIC assembly installed on a chipboard, and is identified by a serial number.
Throughout the data-collection period, the first two slots (N35, N36) of the upstream arm and the last two slots (N10, N161 or N38) of the downstream arm were instrumented with $100~\mum$ thick planar \nonp sensors.
The remaining four slots closest to the DUT in the two arms were equipped with $300~\mum$ thick planar \nonp sensors (N18, N22, N33, N34).
The Timepix4 ASIC plus sensor assemblies are mounted on chipboards that are part of the SPIDR4 readout system\footnote{For more information see \url{http://spidr4.nikhef.nl.}}.
The chipboard hosts low drop-out voltage regulators (LDO) that stabilise the analogue and digital supply voltage close to the Timepix4 chip to minimise fluctuations in the supply voltage. 
Both the digital and analog LDO get their voltage from the same channel of a bench power supply\footnote{Rohde \& Schwarz HMP4040.}, set to 2~V, and placed at 1.5~m distance from the beam.
The chipboard houses sensors for ambient pressure and humidity, and it also has a temperature sensor mounted on the backside of the board opposite to the Timepix4 ASIC.
Each chipboard is connected to a dedicated SPIDR4 Controlboard, which controls and reads out the Timepix4 chip. The controlboard is based on a Xilinx\footnote{Xilinx is part of Advanced Micro Devices, AMD. 2485 Augustine Drive, Santa Clara, CA 95054, United States.} Zync 7000 FPGA, and obtains its reference clock from an external source. 
The connection between chipboard and control board is via a passive feedthrough board (shown in red in \cref{fig:telescope}), which passes signals to the light-tight and gas-tight enclosures.
The communication between the control board and run-control computer is by gRPC, an open source remote procedure call framework\footnote{For more information see \url{https://grpc.io.}}.

\subsection{Data acquisition} 
\label{sec:daq} 
The data from the 2.56 Gbps serial links of Timepix4, one for each side of the chip, are merged by the FPGA on the Control Board and collected into 10-Gbit ethernet packets, which are sent via an optical point-to-point UDP connection to a DAQ computer. 
The SPIDR4 controlboards, and subsequently the Timepix4 ASICs, are time synchronised by a common clock system, consisting of a HPTC 40.08 MHz clock\footnote{GitLab for HPTC Project, CERN. https://gitlab.cern.ch/HPTD/hptc.}, and an LMK1612 differential LVDS clock fanout board, which distributes the clock signal to all SPIDR4 boards. 
The PicoTDC\cite{picotdc}
is also synchronised to the same clock. 
This clock distribution tree provides a system clock with a time resolution better than 15 ps as verified by measuring the edge-to-edge clock jitter between two planes\footnote{Timepix4 can be configured to output its system clock.}.

The PicoTDC is read out via 1-Gbit ethernet using a Xilinx VC707 development board. The firmware and software is based on the test system provided by the PicoTDC group, but was modified to provide an extended timestamp (by 28 bits) and to improve the data throughput.

At the start of each run, the Timepix4 time counters are reset by the Telescope Logic Unit (TLU), which provides an LVDS level pulse (T0-sync), synchronised to the system clock. 
The TLU sends this T0-sync in parallel to all SPIDR4 systems, which in turn pass it to the Timepix4 chip. 
To start and stop the data taking synchronously, the TLU provides a clock-synchronised shutter signal, via the same HDMI cable as the T0-sync.
The TLU is controlled by the run-control computer, and receives the commands to send the T0-sync pulse and to control the state of the shutter via a 1~Gbit ethernet interface.
The SPIDR system is described in ref.~\cite{Akiba:2022fnt}.

The telescope measurement data are recorded by dedicated DAQ servers, each equipped with temporary storage consisting of two \SI{16}{\tera\byte} disks in a RAID~1 setup to protect against disk failure. 
During operation new measurement data is continually transferred to the EOS service at CERN~\cite{Peters:2015aba} for permanent storage. There are four DAQ servers reserved for the telescope planes (two planes on each server) and one server for the DUT. 
Each server runs two instances of the SPIDR4 DAQ application, recording all ethernet UDP datagrams to disk in an uncompressed binary format. 
The run-control computer controls the DAQ application via a gRPC interface through which it also provides metadata that is stored in the data files.

\subsection{Time reference system}
The telescope is equipped with a time reference system to assess the timing performance of the telescope itself and enable tests of new sensors and ASICs with a better time precision than the Timepix4 ASIC.
The time-reference system is composed of two microchannel plate photomultipliers (MCPs) placed in the beam downstream of the telescope.
The MCP closer to the telescope is a Photonis PP2365Y\footnote{Photonis Scientific Inc, 660 Main Street, Fiskdale MA, Sturbridge, MA 01566, United States.} while the more downstream device is a Photek PMT240\footnote{Photek, 26 Castleham Road, St. Leonards on Sea, East Sussex, TN38 9NS, United Kingdom.}.
Both MCPs have a circular window of quartz with a diameter of $16.2$ and $40\mm$,  and a thickness of $5.5$ and $9\mm$, respectively.
The PMT240 MCP is operated at $3.6\kv$, while the PP2365Y MCP is operated at $2.45\kv$.
The resolution of the two MCPs is determined to be less than $5\ps$ based on an oscilloscope measurement of the amplitudes produced by particles from the SPS beam.
The timestamps from the MCPs are refined by two Ortec 9327\footnote{Advanced Measurement Technology, Spectrum House, 1 Millars Business Centre, Fishponds Close, Wokingham, Berkshire, RG41 2TZ, United Kingdom.} 
amplifiers and constant fraction discriminators (CFDs), and they are acquired via a PicoTDC time-to-digital converter (TDC) board\footnote{CERN, Esplanade des Particules 1, P.O. Box, 1211 Geneva 23, Switzerland.} 
operated with $12\ps$ time binning.

\begin{figure}[tb]
    \centering
    \includegraphics[width=0.4\textwidth]{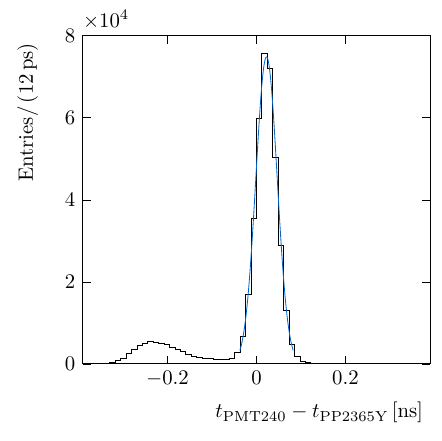}
    \vspace{-2mm}
    \caption{Difference between the track timestamps as obtained from the two MCPs and digitised by the PicoTDC board. The fit of a Gaussian function to the core of the distribution is overlaid.
    The second peak of the distribution is due to energy deposits from multiple tracks coming from hadronic showers, as explained in the text.}
    \label{fig:mcp}
\end{figure}

The difference between the digitised time of the two MCP signals, as read out from the PicoTDC board, is shown in \cref{fig:mcp}.
The lower and wider component of the distribution on the left of the Gaussian core of the distribution is due to energy deposits from particle showers, mostly originating from hadronic interactions in the 
upstream MCP.
The particles originating from these showers deposit more Cherenkov light in the window of the downstream MCP.
As a result, the signal amplitude is outside the operating range of the CFD ($-[15,150]\mv$), which is tuned for operation with single-particle production of Cherenkov light, and hence the CFD is unable to correct for timewalk.
Such events can be vetoed by setting a fiducial window on the time difference, shown in \cref{fig:mcp}, owing to the clear separation of the core of the distribution from the left tail. 

The time measurements of the MCPs are complemented by three plastic (EJ200\footnote{ELJEN technology, 1300 W. Broadway, Sweetwater, TX 79556, United States.}) scintillators.
Two are mounted onto the box of the first telescope arm, upstream of the pixel sensors and spaced approximately $2\cm$ apart from each other, while the third is placed downstream of the second telescope arm. 
The scintillators are instrumented with HPK\footnote{Hamamatsu Photonics K.K., 325-6, Sunayama-cho, Naka-ku, Hamamatsu City, Shizuoka Pref., 430-8587, Japan.} Photomultiplier Tubes (PMTs) and their signals are processed by ORTEC-584 CFDs to minimise the contribution of timewalk to the time resolution.
The discriminated MCP and scintillator signals are fed into the PicoTDC board and also to different Timepix4 planes, where they are timestamped with the TDC of one of the pixels.
The individual scintillators are determined to have a resolution of 120 to 150\ps, depending on the scintillator. This includes the CFD contribution, and is used to verify the synchronisation of the PicoTDC board and Timepix4 ASICs.

\subsection{Device under test}
The main aim of the telescope is to study in detail sensors placed in the DUT position, at the centre of the two telescope arms.
The telescope is able to provide the intercept position of a track with excellent spatial resolution, as reported in \cref{sec:pointingResolution}.
In addition, the time is recorded by the time reference system and by each of the telescope planes for comprehensive timing studies. 
The timestamp created by the combination of the Timepix4 measurements composes the track time and its resolution is reported in \cref{sec:tracktimeresolution}. 
A high rate of tracks can be reconstructed by the telescope, enabling high statistical precision in a short period of time.
In conjunction with the 4D pointing measurements, the telescope is capable of rapidly providing the necessary data for conducting intrapixel analyses.

The DUT area is equipped with a hermetic enclosure featuring thin aluminised Mylar windows. 
This enclosure is mounted on multiple positioning stages that enable $xy$ translations, rotation about the $y$ axis (rotation $\phi$), and tilting about the $x$ axis of the DUT (rotation $\theta$). 
The bidirectional repeatability of these stages is $8\mum$ for the $x$ and $y$ translation, 0.6~mrad for $\varphi$ and 1.1~mrad for $\theta$.
The relative position along the $z$ axis can be adjusted by repositioning the telescope arms.
The DUT enclosure is designed to accommodate silicon assemblies and is already prepared to operate with a Timepix4 device with all the necessary electrical and cooling feedthroughs similar to those of the telescope planes.

\subsection{Experiment control system} 
\label{sec:experimentControlSystem} 
\begin{figure}[tb]
	\centering
	\includegraphics[width=\textwidth]{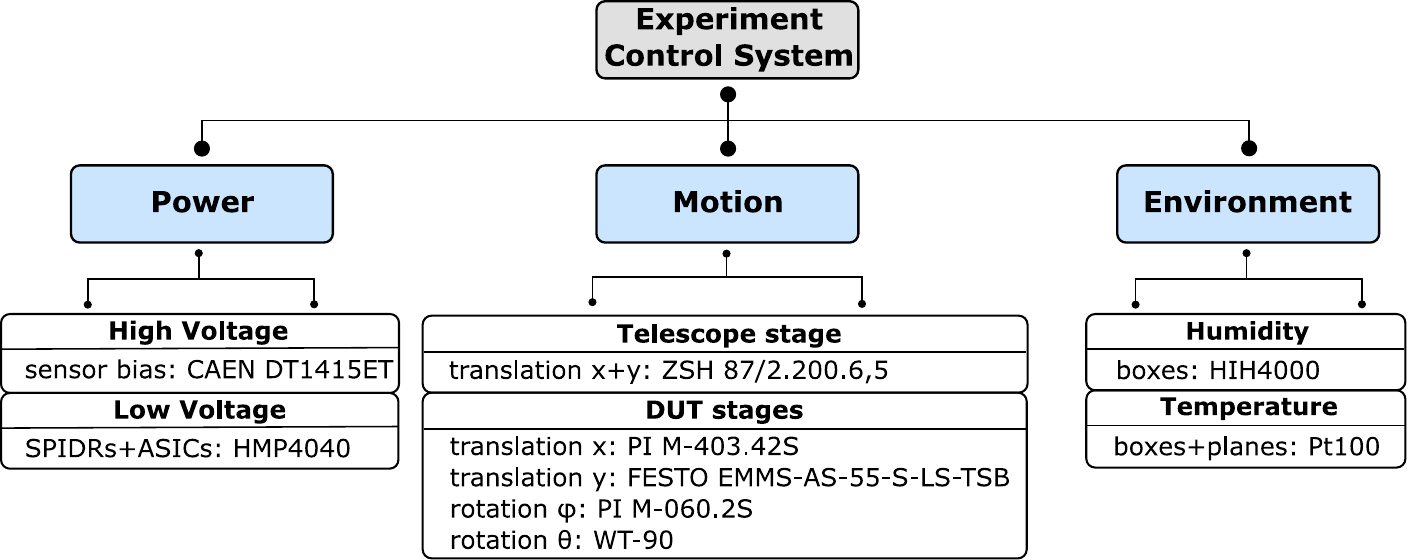}
	\caption{ Schematic overview of the ECS components and hardware. }
	\label{fig:tpx_ecs}
\end{figure}

The telescope is remotely operated and monitored by the Experiment Control System (ECS). 
It controls the high and low-voltage power supplies, operates the telescope and DUT motion stages, and monitors the temperature and humidity of the environment. 
Details of the implementation are described in ref.~\cite{Akiba:2022fnt}.
A block diagram of the system is shown in \cref{fig:tpx_ecs}.

The sensor planes of the telescope are biased by an eight channel CAEN DT1415ET\footnote{CAEN SpA, Via Vetraia 11, 55049 Viareggio (LU), Italy.} providing high voltage to each of the planes.
The DUTs are biased independently by Keithley 2410 Source Meters\footnote{Tektronix, Inc. 14150 SW Karl Braun Drive, Beaverton, OR 97077, United States.}. 
The low voltage for the SPIDR4 boards and the ASICs is provided by two HMP4040 programmable power supplies\footnote{Rohde \& Schwarz USA, Inc. 6821 Benjamin Franklin Drive, Columbia, MD 21046, United States.}. 
The motion of the telescope is performed with precision translation and rotation stages from PI\footnote{Physik Instrumente (PI) GmbH \& Co. KG, Auf der Roemerstrasse 1, 76228 Karlsruhe, Germany.}, FESTO\footnote{Festo Corporation 1377 Motor Parkway, Islandia, NY 11749, United States.}, and Phytron\footnote{Phytron GmbH, August-Rasch-Straße 11, D-82216 Maisach, Germany.}. 
The temperatures of each plane, as well as the temperature and humidity within the hermetic enclosures, are monitored with four-wire Pt100 (attached to the cooling block) and HIH4000 sensors\footnote{Honeywell International Inc. 855 South Mint Street, Charlotte, NC 28202, United States.}. 

\subsection{Reconstruction software and data quality monitoring}
The event reconstruction is based on the \kepler software package, as described in ref.~\cite{Akiba:2019faz}, updated to be able to decode the Timepix4 and PicoTDC data packets.
A list of the standard selection requirements for the event reconstruction is given in \cref{tab:selection_requirements}.
A graphical user interface, implemented using the Qt5 toolkit and communicating with the \kepler server through the DIM protocol~\cite{Gaspar:2001fbw}, is used to monitor the quality of the collected data in real time.
The monitored information consists mostly of histograms of quantities such as spatial and ToT distributions, as well as properties related to the clusters or tracks.

\begin{table}[]
    \centering
    \caption{List of the standard selection requirements for the track reconstruction.}
    \label{tab:selection_requirements}
    {\footnotesize 
    \begin{tabularx}{\textwidth}{llX}
        \toprule
         Requirement & Default value & Description \\ \midrule
         Cluster time window & $ 100 \ns$ & Maximum time difference of hits within the same cluster \\ 
         Cluster width in $x$ and $y$ & $\leq 2$ & Rejects large clusters from $\delta$-rays, nuclear interactions, etc. \\ 
         Track time window & $5 \ns$ & Clusters within $5 \ns$ are considered for the pattern recognition \\ 
         Number of clusters in time window & $ 1 \, \textrm{per plane}$ & Rejects multiprong interaction vertices \\ 
         Number of clusters per track & $8$ & Maximizes track precision \\ 
         Opening angle & $< $ \SI{10}{\mrad}  & Angle that defines the reconstruction window from plane to plane, assuming straight tracks \\ 
         Fit $\chi^2 / \mathrm{ndof}$ & $< 10$ & Cut on $\chi^2$ divided by the number of degrees of freedom for track quality \\ \bottomrule
    \end{tabularx}
    }
\end{table}


\section{Measurement calibrations} 
\label{sec:corrections} 
\subsection{Charge calibration}
\label{sec:chargecalib}
The analogue front-end of Timepix4 uses a charge-sensitive amplifier (CSA) with a feedback capacitor for amplification of the input signal.
The input signal current is integrated onto the feedback capacitor, which is then discharged via a constant current controlled by a Krummenacher scheme~\cite{Krummenacher:1991qhr}. 
The CSA output is connected to a leading-edge threshold discriminator. 

The length of the output pulse of the CSA, and hence to first order also the ToT, is proportional to the amount of charge because the CSA discharge current is constant.
However, for low charge, the ToT does not scale linearly with the CSA output pulse length, because the switching off of the discriminator is slow, in particular for low thresholds.

\begin{figure}[tb]
    \centering
        \begin{subfigure}{0.495\textwidth}
    \includegraphics[width=\textwidth]{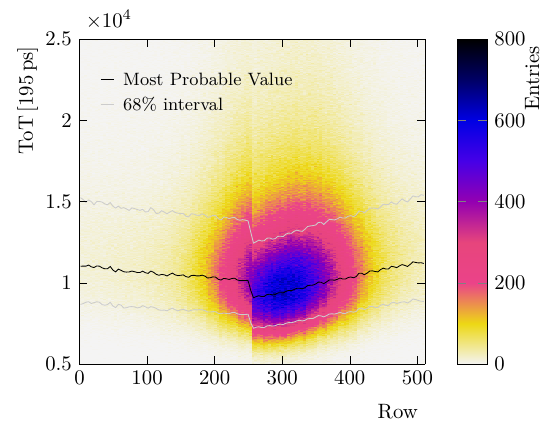}
    \vspace{-2\baselineskip} 
    \subcaption{} \label{fig:tot_before_correction:a}
    \end{subfigure}
    \begin{subfigure}{0.495\textwidth}
    \includegraphics[width=\textwidth]{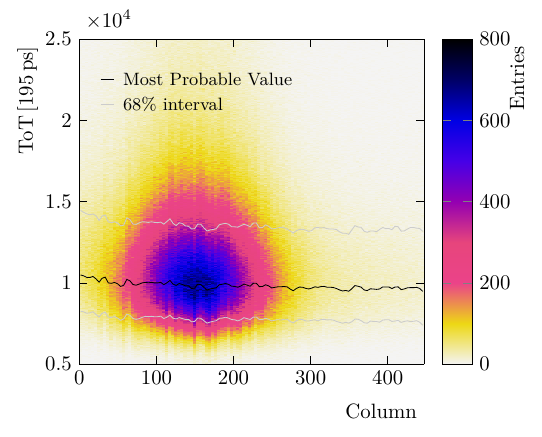}
    \vspace{-2\baselineskip} 
    \subcaption{} \label{fig:tot_before_correction:b}
    \end{subfigure}
    \caption{ToT as a function of row (a) and column (b) for single pixel clusters in a $100\mum$ telescope plane. 
    The solid line shows the most probable value of ToT in the given row or column and the grey lines the central $68\%$ interval. }
    \label{fig:tot_before_correction}
\end{figure}

The ToT exhibits a row dependence towards the centre of the matrix, with a discontinuity in the middle, as shown in \cref{fig:tot_before_correction:a}. 
The row dependence is due to a gradual supply voltage drop towards the centre because power is supplied from the top and bottom only. 
The Timepix4 is compatible with through silicon vias, which allow supply voltage connections also to the centre of the chip. 
This will reduce the voltage drop, hence the ToT variation by a factor of four. 
However, the Timepix4 ASICs of the current telescope do not use this feature. 
The discontinuity in the middle is caused by a systematic difference in the DAC voltages for the top and bottom halves of the pixel matrix.
As shown in \cref{fig:tot_before_correction:b}, the ToT also exhibits vertical banding across the columns. 

\begin{figure}[tb]
    \centering
    \includegraphics[width=0.4\textwidth]{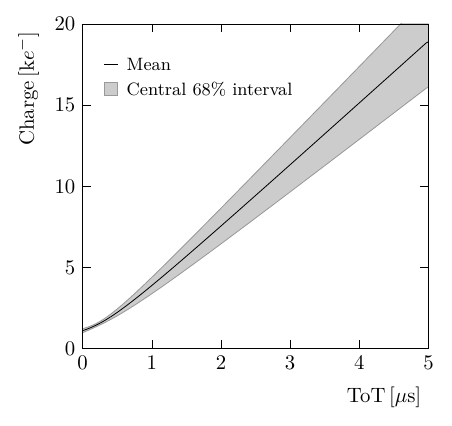}
    \vspace{-2mm}
    \caption{Charge distribution for all pixels in plane N10 as a function of ToT. 
    The solid line shows the mean value of charge for the given ToT and the shaded band the central $68\%$ interval.} 
    \label{fig:tot_to_charge}
\end{figure}

A per-pixel charge calibration is performed to convert the measured ToT to charge.
The calibration procedure injects test pulses with a controlled amount of charge into the analogue frontend, and measures the corresponding ToT.
A fit is performed to the ToT response using the surrogate function
\begin{equation}
    \textnormal{ToT}(q) = p_0 + p_1 q - \frac{p_2}{q - p_3},
    \label{eqn:surrfunc}
\end{equation}
where $p_0$ corresponds to an offset, $p_1$ the conversion gain, $p_2$ and $p_3$ describe the nonlinear response at low charge \cite{Jakubek:2011dsm}. 
The test pulses are injected in 0.2~\ke (0.5~\ke) increments, ranging from 0.0 to 4.0~\ke (4.0 to 20.0~\ke).
Due to limitations in the power supply of the test pulse circuitry, only one pixel in a sub-matrix of at least 64 pixels receives the injected charge at a given step. 
The resulting data are fitted using \cref{eqn:surrfunc}. 
For sample points to be used in the fit, at least 45 of the test pulses must have been registered. 
The inverse of this function is used to convert from ToT to charge. \Cref{fig:tot_to_charge} shows the spread in the relationship between ToT and charge for a single Timepix4 plane. 
The wide distribution (in ToT) is due to ToT-gain variation across the chip. 

\begin{figure}[tb]
    \centering
        \hspace{10mm} 
        \includegraphics[width=0.495\textwidth]{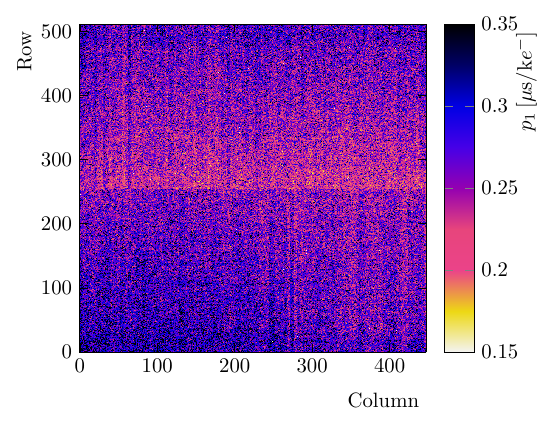}
    \caption{Variation of the conversion gain, $p_1$, across the chip in plane N10.}
    \label{fig:p1_var}
\end{figure}

The per-pixel variation of gain across a single module is shown in \cref{fig:p1_var}, where the structure shown in \cref{fig:tot_before_correction} can be seen more easily. 

\begin{figure}
    \centering
        \begin{subfigure}{0.495\textwidth}
    \includegraphics[width=\textwidth]{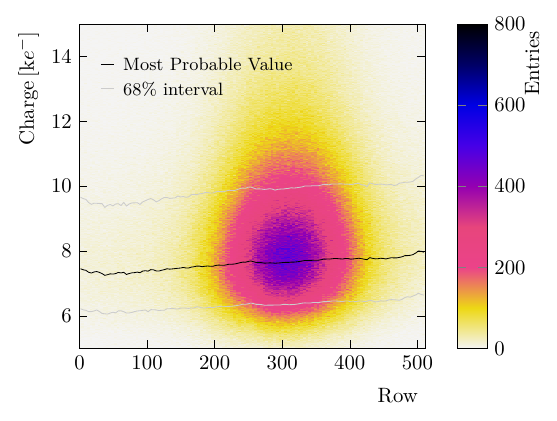}
        \vspace{-2\baselineskip} 
    \subcaption{} \label{fig:charge_after_correction:a}
    \end{subfigure}
        \begin{subfigure}{0.4815\textwidth}
    \includegraphics[width=\textwidth]{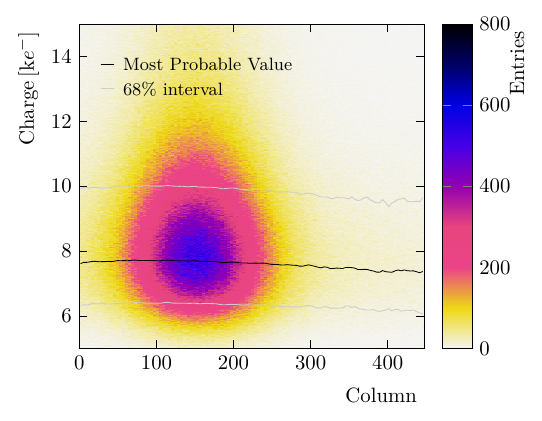}
    \vspace{-2\baselineskip} 
    \subcaption{} \label{fig:charge_after_correction:b}
    \end{subfigure}
    \vspace{-2mm}
    \caption{Calculated charge as a function of row (a) and column (b) for single pixel clusters in a $100~\mum$ telescope plane. 
    The solid black line shows the most probable value of the charge in the given row (column) and the grey line indicates the central $68\%$ interval. }
    \label{fig:charge_after_correction}
\end{figure}

The large variations in ToT conversion gain across the chip are effectively mitigated by the calibration, as can be seen in ~\cref{fig:charge_after_correction}.
The most probable value of deposited charge integrated over all pixels in the timing (spatial) planes is around 7.4~\ke (23.5~\ke), roughly in line with the expected value for 100~\mum (300~\mum) thick silicon sensors \cite{Workman:2022ynf}.

Charge calibration using test pulses, as performed here, has a systematic error proportional to the amount of charge being measured. 
There is an uncertainty in the test pulse capacitance due to process variations during the fabrication, leading to a proportional error in the amount of charge inserted into the frontend. 
By performing ${}^{55}\textnormal{Fe}$ and ${}^{109}\textnormal{Cd}$ measurements, it has been confirmed that there is a significant difference in the charge calibration. 
This effect has also been previously reported~\cite{Delogu:2024}. 
Radioactive-source measurements also indicate that the proportional error in the charge calibration does not vary significantly over the pixel matrix of a single Timepix4 plane. 
Therefore, the charge calibration acquired with test pulses is suitable for the track reconstruction since the charge-weighted cluster position is based on fractional values, and thus only requires a uniform proportionality to ToT across the matrix.

\subsection{Corrections to spatial  measurements}
\label{sec:etacorr}
The charge-weighted average of the hits in a cluster is a biased estimator of the true position at which a particle traverses the sensor. 
At small incident angles, the bias is due to the induced signal on adjacent pixels mostly coming from the lateral diffusion of the charge. 
Conversely, at larger angles more of the signal in subleading pixels is from charge liberated within that pixel, resulting in a smaller bias.

\begin{figure}
    \centering
    \begin{subfigure}{0.495\textwidth}
    \includegraphics[width=\textwidth]{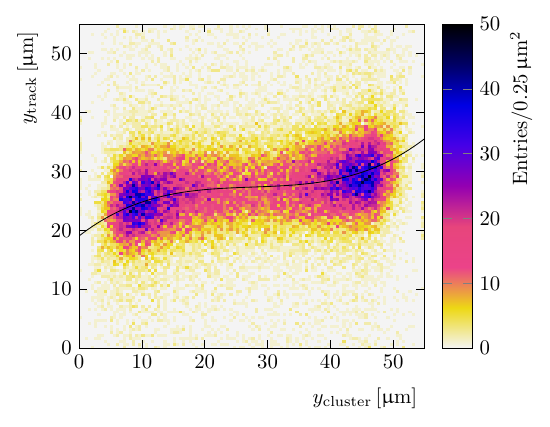}
    \vspace{-2.0\baselineskip} \subcaption{} \label{fig:charge_sharing_corrections:a}
    \end{subfigure}
    \begin{subfigure}{0.495\textwidth}
    \includegraphics[width=\textwidth]{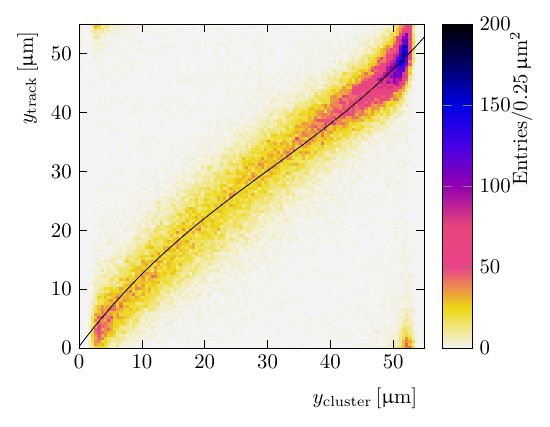}
    \vspace{-2.0\baselineskip} \subcaption{} \label{fig:charge_sharing_corrections:b}
    \end{subfigure}

    \begin{subfigure}{0.495\textwidth}
    \includegraphics[width=\textwidth]{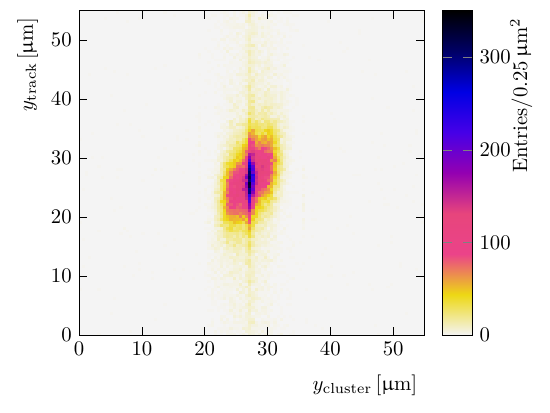}
    \vspace{-2.0\baselineskip} \subcaption{} \label{fig:charge_sharing_corrections:c}
    \end{subfigure}
    \begin{subfigure}{0.495\textwidth}
    \includegraphics[width=\textwidth]{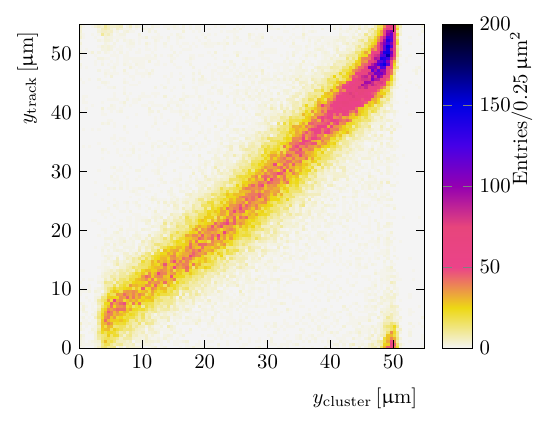}
    \vspace{-2.0\baselineskip} \subcaption{} \label{fig:charge_sharing_corrections:d}
    \end{subfigure}
    \vspace{-2mm}
    \caption{Track position compared to charge weighted cluster position before (a,b) and after (c,d) correction for two-pixel clusters in a perpendicular (a,c) and inclined (b,d) telescope plane. 
    The third-order polynomial used for correction is overlaid on the initial distribution. A charge-weighted cluster position of 27.5~\mum represents the boundary between two adjacent pixels. }
    \label{fig:charge_sharing_corrections}
\end{figure}

The track position is compared to the charge-weighted position for associated two-pixel clusters in \cref{fig:charge_sharing_corrections},  for both perpendicular and inclined planes.
Charge sharing in the inclined planes yields a more linear relationship as the inclined geometry gives rise to direct charge deposition in multiple pixels.
Following the method in ref.~\cite{Akiba:2011vn}, a fit is performed to the average of the position, as estimated by the charge-weighted average, as a function of intrapixel position with a third-order polynomial. 
The fit is used to correct residual non-linearities, known as $\eta$ corrections.
Since the track position used in this procedure depends on the telescope alignment, it is repeated iteratively, with the $\eta$-corrected cluster positions used during the track fit in the alignment process. 
The iteration is repeated until no further improvement to the resolution is found. 
Typically, this happens after two iterations.
The corrected histograms from the final iteration are also shown in \cref{fig:charge_sharing_corrections}.
The correction calculated from two-pixel clusters in both the $x$- and $y$-directions is applied to all clusters that contain hits in at most two rows and two columns. 
The corrections are found to improve the cluster spatial resolution of the thin (thick) planes by about 8\% (21\%).

\subsection{Spatial alignment}
\label{sec:alignment}
\begin{figure}
  \centering
  \begin{subfigure}{0.45\textwidth}
  \includegraphics[width=\textwidth]{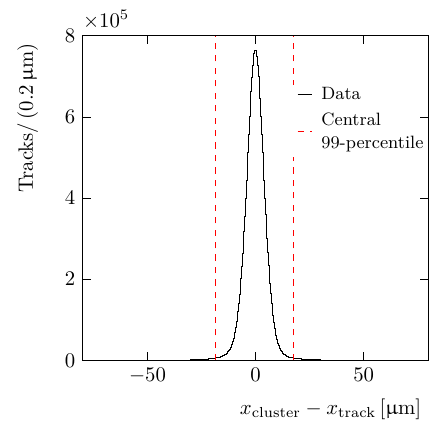}
  \vspace{-2\baselineskip}
  \caption{}
  \label{fig:residuals_thick:a}
  \end{subfigure}
  \begin{subfigure}{0.54\textwidth}
  \includegraphics[width=\textwidth]{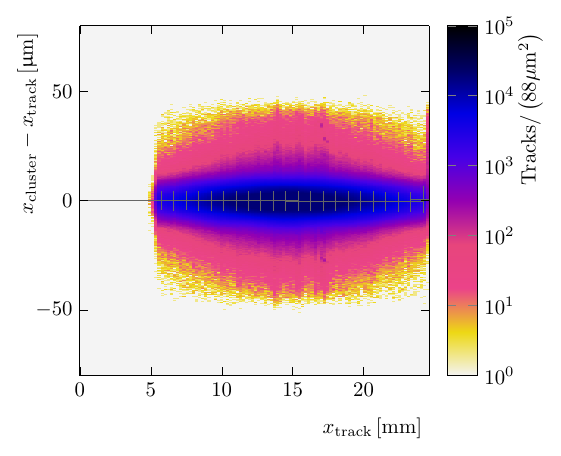}
  \vspace{-2\baselineskip}
  \caption{}
  \label{fig:residuals_thick:b}
  \end{subfigure}

  \begin{subfigure}{0.45\textwidth}
  \includegraphics[width=\textwidth]{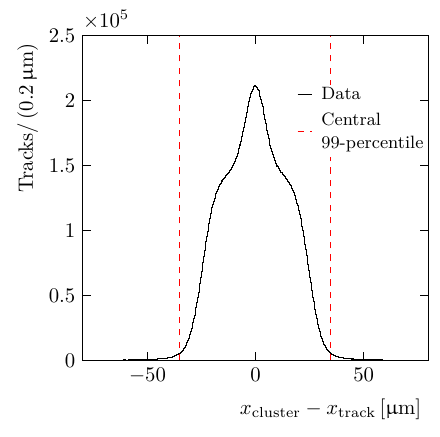}
  \vspace{-2\baselineskip}
  \caption{} \label{fig:residuals_thick:c}
  \end{subfigure}
  \begin{subfigure}{0.54\textwidth}
  \includegraphics[width=\textwidth]{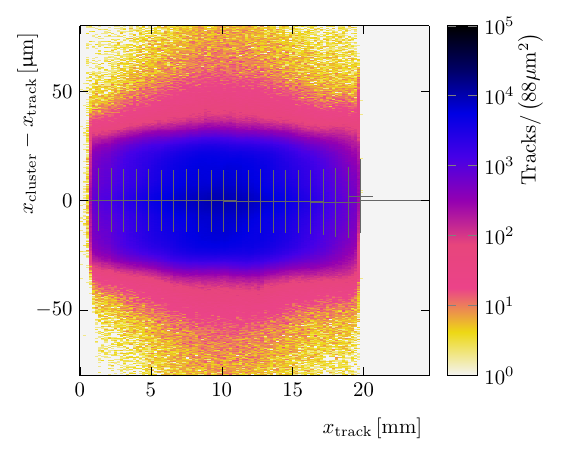}
  \vspace{-2\baselineskip}
  \caption{} \label{fig:residuals_thick:d}
  \end{subfigure}
  
  \caption{
      Unbiased residuals in $x$ for one thick (a,b) and one thin (c,d) plane, integrated over the whole (a,c) sensor and as a function of the local position (b,d).
      The error bars show the fitted width at each position.
  }
  \label{fig:residuals_thick}
\end{figure}

The spatial alignment of the telescope is performed using the \textsc{Millepede} \cite{Blobel:2006yh} algorithm implemented in \kepler.
The precision of the spatial alignment of each plane is measured in an unbiased manner, by excluding the plane from the tracking algorithm and comparing the extrapolated track position with the position of the corresponding cluster on that plane.
These unbiased residuals, integrated throughout the sensor and as a function of the local position, are displayed in \cref{fig:residuals_thick}. 
The position-dependent residuals can be used to check for rotations around the $x$ and $y$ axes.
The unbiased residual distributions exhibit a smaller spread for the thick planes due to their tilt with respect to the beam axis, which improves the spatial resolution of these planes.
The mean and standard deviation are determined for the central 99\,\% of the distribution.
The variation of the average residual across the beam spot relative to all axes is found to be less than $1\mum$ on all planes, indicating a good alignment.

The systematic error on the spatial resolution of each plane is estimated by analysing the variations of the standard deviation of the residual distributions across the beam spot in both $x$ and $y$.
A spread of 1.1\% to 6.2\% is obtained for the different planes.

\subsection{Corrections to timing measurements}
\label{sec:timecorrections}
The time measurements are corrected to compensate for effects that degrade the resolution.
Firstly, a time alignment is performed consisting of the subtraction of the average time difference between the pixel hits and the time reference. 
The time reference is given by the average of the two MCP measurements. 
Secondly, nonuniformities arising from differences in the VCO frequency between superpixels are compensated by estimating the actual frequency for each pixel based on particle data, as described in \cref{sec:vco_correction}. 
The correction of timewalk, which is the dependence of the time of arrival on the amplitude of the signal is described in \cref{sec:timewalkCorrection}.
The corrections are applied as 
\begin{equation}
    t_{\mathrm{corrected}}(\mathrm{row, col}, q) = \mathrm{ToA}_{\mathrm{VCO\;corrected}}(\mathrm{row, col}) - \Delta t_{\mathrm{align}} - \Delta t_{\mathrm{timewalk}}(\mathrm{row,col} , q),
    \label{eq:time:corrected_time}
\end{equation}
where $\mathrm{ToA}_{\mathrm{VCO\;corrected}} $ is the measured time of arrival taking into account the real VCO frequency.

The temporal performance of each plane is evaluated using the time residuals with respect to the reference time. 
The resolution, $\sigma_t$, is defined as the standard deviation of the central 99\% of the obtained distribution. 

\subsubsection{Time alignment}
\label{sec:time_alignment}
Each telescope plane is time aligned by subtracting a time offset to compensate for cable delays and differences in propagation delay in the electronics.
The time offset is determined as the average time difference for each plane with respect to the reference time.

\subsubsection{VCO Correction}
\label{sec:vco_correction}
As described in \cref{sec:timepix4}, each superpixel is equipped with its own VCO, to measure the arrival time of hit using \cref{eq:time:corrected_time}.
In an ideal ASIC, all VCOs would oscillate at a frequency of  640~MHz. 
However, variations in the process parameters of the VCO components, such as transistors and capacitors, along with differences in local supply voltage and non-uniform temperature across the chip, lead to a frequency spread across the matrix.
While frequency differences between superpixels are expected, 
the pixels within a superpixel have the same frequency as they share a VCO. 
The observed spread in the VCO frequencies between pixels is 0.7\% for all planes within the region covered by the beam, with 99\% of measured pixel frequencies within $\pm$13~MHz of the nominal value.

Since it is not possible to measure the frequency of each VCO directly, a statistical method is used to extract the frequencies from beam data. 
If the VCO in a superpixel oscillates exactly at 640~MHz, and hits arrive at random times with respect to the 40~MHz clock, the distribution of fine Time-of-Arrival (fToA) values would have 16 bins with equal counts (within statistical error). 
However, due to the specific implementation of the VCO circuit, the histogram of a perfect VCO has 17 bins, where the first and last bins contain half the number of counts compared to the other bins. 
If the VCO runs slightly faster (or slower) than 640~MHz, the number of counts in the last bin will be higher (or lower). For significantly higher frequencies, even the 18$^\text{th}$ bin may contain counts. 
The frequency of the VCO, $f_{\mathrm{VCO}}$, can be determined from
\begin{equation}
    f_\text{VCO} = \left( \frac{n_0 + \sum_{i = 16}^{31}n_{i}  }{\overline n_{1-15}} + 15 \right) \times \frac{\SI{640}{\mega\hertz}}{16} 
    \label{eq:time:frequency}
\end{equation}
where  $n_i$ is the count in the $i^{th}$ bin and $\overline n_{1-15}$ is the average count in bins 1 to 15.

\begin{figure}
    \centering
    \includegraphics[width=0.495\textwidth]{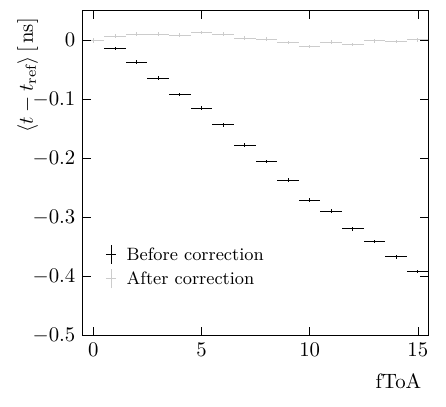}
    \caption{
          The average difference between measured and reference time before and after the VCO correction.
    }
    \label{fig:time:VCO_correction}

\end{figure}

The deviation of the VCO frequency is shown in \cref{fig:time:VCO_correction}, where the average difference between the measured time, $t$, and the reference time, $t_\text{ref}$, is plotted as a function of the fToA bin. 
The discrepancy increases for higher fToA bin values because of the difference in actual and assumed VCO frequency.  
The accumulated bias is removed, as shown in \cref{fig:time:VCO_correction}, by applying the corrected VCO frequency. 

\subsubsection{Timewalk Correction}
\label{sec:timewalkCorrection}
\begin{figure}  
  \centering
	  \begin{subfigure}{0.495\textwidth}
	  \includegraphics[width=\textwidth]{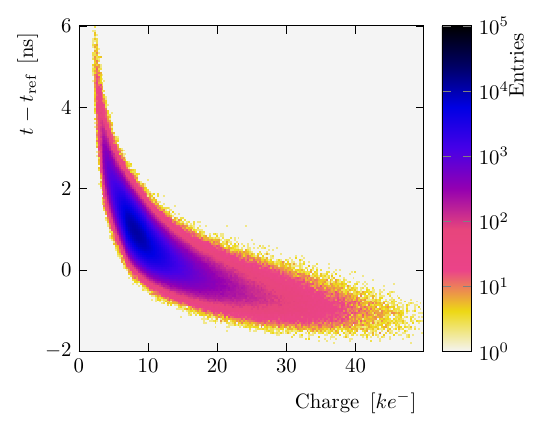}
	  \vspace{-2.0\baselineskip}
	  \subcaption{}\label{fig:time:timewalk_curve:a}
	  \end{subfigure}
	  \begin{subfigure}{0.495\textwidth}
	  \includegraphics[width=\textwidth]{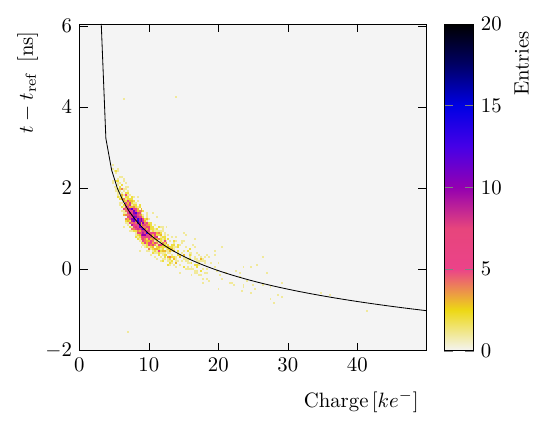}
	  \vspace{-2.0\baselineskip}
	  \subcaption{}\label{fig:time:timewalk_curve:b}
	  \end{subfigure}
  
  \caption{
   Distribution of time residuals as a function of charge for all pixels for plane N161 ($100~\mum$) (a) and a single pixel with its corresponding fit overlaid (b).
  }
  \label{fig:time:timewalk_curve}
\end{figure}

Timewalk refers to the variation as a function of charge in the measured time of signals when using fixed-threshold discrimination, and can be seen in \cref{fig:time:timewalk_curve:a} for plane N161 ($100~\mum$). 
This effect occurs due to fluctuations in signal amplitude combined with a non-zero rise time, which causes the detection threshold to be crossed at different times relative to the true time-of-arrival.
In the results shown in this paper, the timewalk measured reflects a combination of the effects caused by the discriminator and the finite speed of charge carriers within the sensor.

\begin{figure}  
  \centering
  \includegraphics[width=0.495\textwidth]{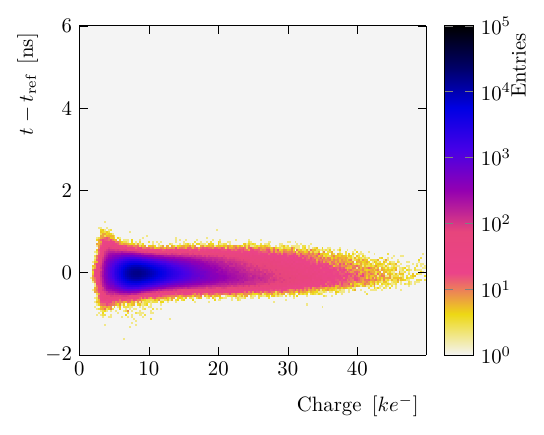}
  \caption{The distribution of time residuals as a function of charge, after correcting for timewalk, for all pixels in plane N161 (100\mum).
  }
  \label{fig:timewalk_corrected}
\end{figure}

A timewalk correction as implemented for Timepix4 ASICs has been previously described \cite{Akiba:2022fnt}. 
In this analysis, an improved correction is implemented using ToT converted to charge as described in \cref{sec:chargecalib}, compensating for conversion gain fluctuations across the ASIC.
The timewalk correction is modelled as a function of the 
charge ($q$) using the empirical formula
\begin{equation}
    \Delta t_{timewalk}(q) = \frac{a}{(q + b)^c}+d,
    \label{func:timewalk}
\end{equation}
where  $a$, $b$, $c$ and $d$ are the parameters of the fit and are determined for each pixel.
In comparison to the method employed in ref.~\cite{Akiba:2022fnt}, this function has an additional parameter ($c$) which improves stability and convergence of the fits. 
The timewalk distribution for a single pixel of sensor N161 ($100 ~\mum$) is shown in \cref{fig:time:timewalk_curve:b}, where the curve is given by \cref{func:timewalk}, with parameters obtained by performing a fit to the distribution. 
The time offset, $d$, has a variation of up to 1.2~\ns ($99\%$ interval), which is considerable with respect to the temporal resolution. 
The distribution of the time residual as a function of charge after applying the timewalk correction is shown in \cref{fig:timewalk_corrected}.

\begin{figure}  
  \centering
  \begin{subfigure}{0.495\textwidth}
  \includegraphics[width=\textwidth]{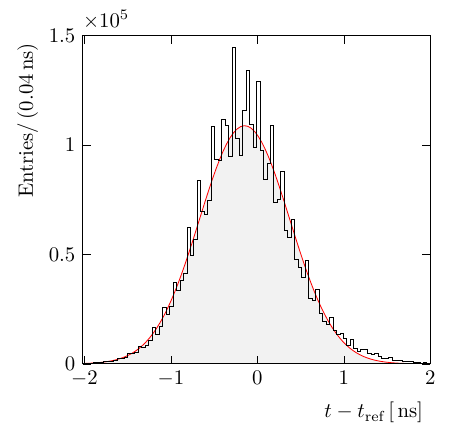}
  \vspace{-2.0\baselineskip}
  \subcaption{}\label{fig:time:vco_correction_resolution:a}
  \end{subfigure}
  \begin{subfigure}{0.495\textwidth}
  \includegraphics[width=\textwidth]{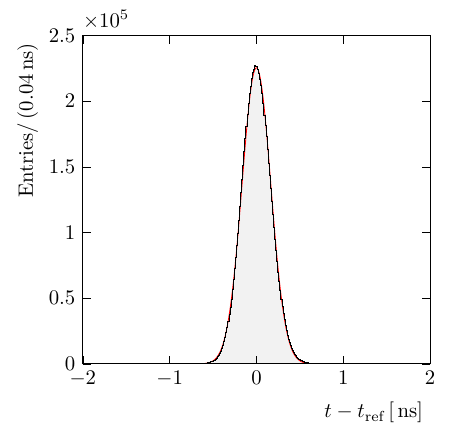}
          \vspace{-2.0\baselineskip}
  \subcaption{}\label{fig:time:vco_correction_resolution:b}
    \end{subfigure}
  \caption{Temporal residual distributions for sensor N161 before any correction (a) and after all corrections (b). On both plots the smooth line represents a Gaussian fit. 
  }
  \label{fig:time:vco_correction_resolution}
\end{figure}

The result of the VCO and timewalk corrections is shown in \Cref{fig:time:vco_correction_resolution}, with the temporal residual shown before and after correction in (a) and (b), respectively.

\section{Results}
\subsection{Spatial resolution}
\label{sec:pointingResolution}
The precision with which the position of the track is known when extrapolated to the position of the DUT is hereafter referred to as pointing resolution.
This resolution depends on the spatial resolution of each plane, on the amount of material traversed by the beam and, to a smaller extent, on the spacing of the telescope planes.
The sensor and ASIC contribute about 0.7--1.0 \% of a radiation length ($X_0$).
The largest material contribution, however, comes from the chipboard, which adds 1.8 to 2.4\% \Xrad of material, 
where the variation comes from the inhomogeneity in the amount of copper.

The pointing resolution is determined through simulation.
An initial estimate of the cluster resolution of the thick and thin planes is obtained from the unbiased residual distributions by subtracting in quadrature a reasonable first guess of the pointing resolution at the $z$-position of each plane.
These resolutions are used in a simulation similar to the one described in ref.~\cite{Akiba:201347}.
The simulation provides the expected biased residual distributions at each plane. These distributions are compared to the data, and the resolutions that are input to the simulation are varied until the best overall agreement is achieved.
The cluster resolutions of the thick and thin planes are found to be \XResThick and \XResThin in $x$ and \YResThick and \YResThin in $y$, respectively.
The output of the simulation is then used to derive the $z$-dependent pointing resolution.
The uncertainty on the pointing resolution is determined by varying the intrinsic spatial resolution of each plane and the material in the beam.

\begin{figure}
  \centering
  \includegraphics[width=0.5\textwidth]{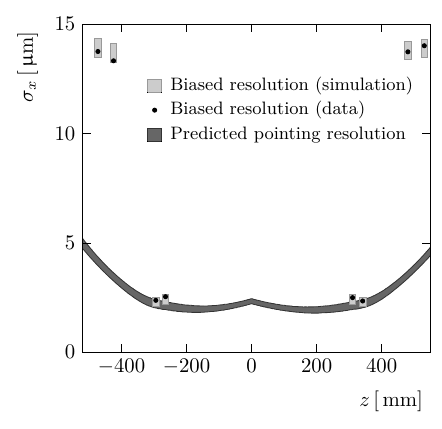}
  \vspace{-\baselineskip}
  \caption{
      Pointing resolution in $x$ as a function of the position along the beam axis, $z$.
      The uncertainty on the pointing resolution is illustrated by the band. The DUT position is at $z = 0$.}
  \label{fig:pointing_resolution}
\end{figure}

The pointing resolution in $x$ as a function of the $z$-position is shown in \cref{fig:pointing_resolution}, and the resolution in the $y$ direction is found to be very similar. 
Pointing resolutions at the DUT position of \XResTrack and \YResTrack are obtained for the $x$ and $y$ directions, respectively.  These results are also summarised in \cref{tab:resolutions}.

\subsection{Temporal performance} 
\label{sec:timePerformance} 
The most precise time measurement of the telescope is provided by the timing reference system instrumented with the two MCPs as described in \cref{sec:setup}. 
However, after the calibration described in \cref{sec:timecorrections}, the time can be provided with excellent precision by each of the telescope planes. 
In addition, the average time per reconstructed track can be obtained, resulting in the track time, which has a better resolution compared to the resolution of each of the individual planes. 
The final track-time resolution is presented in \cref{sec:tracktimeresolution}.

\subsubsection{Single plane temporal resolution}

 The measured time distribution and temporal resolution for a single plane, after applying both VCO and timewalk corrections, is shown in \cref{fig:time:vco_correction_resolution:b}.
From this distribution, a temporal resolution of \SingleTimeRes is obtained for plane 7 (N161), with a bias voltage of -70~V,  at a threshold of 1000~$e^-$.
The obtained temporal resolutions for each plane are summarised in \cref{tab:resolutions}, where all the uncertainties are estimated as the standard deviation of the resolutions determined for several runs under the same conditions.

\begin{table}[]
    \centering
    \caption{ Temporal and spatial resolutions for telescope planes.
    }
    \label{tab:resolutions}
    \begin{tabular}{l
          >{\collectcell\num}r<{\endcollectcell} @{${}\pm{}$} >{\collectcell\num}l<{\endcollectcell}
          >{\collectcell\num}r<{\endcollectcell} @{${}\pm{}$} >{\collectcell\num}l<{\endcollectcell}
          >{\collectcell\num}r<{\endcollectcell} @{${}\pm{}$} >{\collectcell\num}l<{\endcollectcell}
    }
        \toprule
         Plane &  \multicolumn{2}{c}{$\sigma_t$ [\unit{\pico \second}]} & \multicolumn{2}{c}{$\sigma_x$ [\unit{\micro \meter}]} & \multicolumn{2}{c}{$\sigma_y$ [\unit{\micro \meter}]} \\ 
         \midrule 
         N35  & 195 & 13 & 14.4 & 0.5 & 14.3 & 0.5 \\
         N36  & 196 & 14 & 14.4 & 0.5 & 14.3 & 0.5 \\ 
         N33  & 520 & 35 & 3.3  & 0.3 & 3.5  & 0.3 \\ 
         N18  & 605 & 43 & 3.3  & 0.3 & 3.5  & 0.3 \\ 
         N22  & 495 & 21 & 3.3  & 0.3 & 3.5  & 0.3 \\ 
         N34  & 598 & 23 & 3.3  & 0.3 & 3.5  & 0.3 \\ 
         N10  & 170 &  3 & 14.4 & 0.5 & 14.3 & 0.5 \\ 
         N161 & 175 &  5 & 14.4 & 0.5 & 14.3 & 0.5 \\
         \midrule
         Tracks at DUT position & 92 & 5 & 2.3 & 0.1 & 2.4 & 0.1 \\
         \bottomrule
    \end{tabular}
\end{table}

\begin{figure}
  \centering
	  \includegraphics[width=0.495\textwidth]{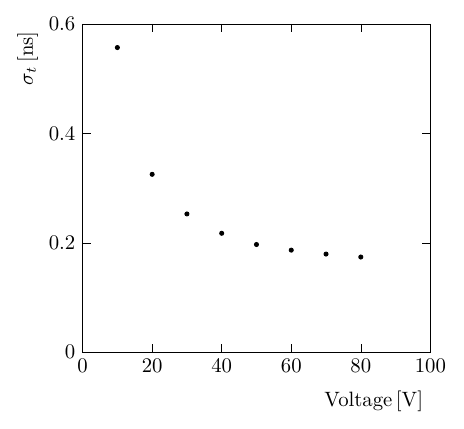}
  \caption{Temporal resolution of the telescope plane N161 (100~\mum) as a function of the bias voltage.
  }
  \label{fig:time:temporalScans}
\end{figure}

The effect of operational bias voltage on the temporal resolution is shown in \cref{fig:time:temporalScans}. 
It is observed across all sensor planes that an increase in bias voltage correlates with enhanced resolution, which can be attributed to the increased velocity of charge carriers,
leading to shorter current pulses from the sensor, resulting in a reduction in jitter.
As the bias is increased further, the improvement in time resolution is modest. However, the velocity of the charge carriers does not yet reach the saturation value and hence it would still be beneficial to use sensors that can tolerate a higher voltage than the currently installed sensors. Tests with a 100~\mum sensor that could be operated at 200~V showed an improvement in the time resolution of about 20~ps.

\subsubsection{Temporal resolution for tracks}
\label{sec:tracktimeresolution}
The track time resolution is determined by calculating a weighted average of the time measurement from all planes in the telescope.
The track time, $t_\text{track}$, is given by  
\begin{equation}
    t_\mathrm{track} =  \frac{\sum_i{{t_i}/{\sigma_i^2}}}{\sum_i {{{1}/{\sigma_i^2}}}},
    \label{eq:tracktime}
\end{equation}
where $t_i$ is the time recorded by each individual plane, and $\sigma_i$ the resolution obtained for each plane, given in \cref{tab:resolutions}. 
The resulting $t_\text{track}$ is used to calculate residuals with respect to the time reference.
The track time resolution is the standard deviation of this residual and is \TrackTimeRes. 

\begin{figure}
  \centering
  {\includegraphics[width=0.495\textwidth]{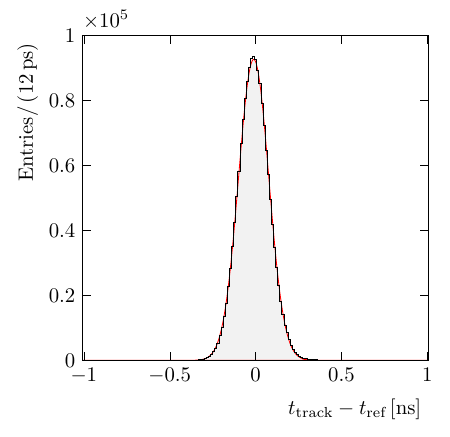}}
  \caption{Track time calculated using all planes fully biased and at a threshold of \mbox{$1000 \, \en$} after both VCO and timewalk corrections. 
  }
  \label{fig:time:TrackRes}
\end{figure}

\subsection{Efficiency and Rate results}
\label{sec:trackingEfficiency}
The single-plane cluster finding efficiency is measured for each telescope plane by excluding the plane from the pattern recognition, reconstructing tracks using the other seven telescope planes and searching for clusters within 165~\mum and 5~ns of the intercept on the plane under investigation. 
The results for the different assemblies are shown in \cref{tab:cluster_efficiencies}.
\begin{table}[]
    \centering
    \caption{Single plane cluster finding efficiencies and the total tracking efficiency of the telescope.
    }
    \label{tab:cluster_efficiencies}
    \begin{tabular}{lll}
        \toprule
         Plane & Thickness $[\mum]$ & Cluster efficiency [\%] \\ %
         \midrule
         N10 & 100 & $98.72 \pm 0.02 $ \\ 
         N18 & 300 & $93.20 \pm 0.20 $ \\ 
         N22 & 300 & $93.37 \pm 0.12 $ \\ 
         N33 & 300 & $93.80 \pm 0.19 $ \\ 
         N34 & 300 & $91.66 \pm 0.17 $ \\ 
         N35 & 100 & $98.52 \pm 0.19 $ \\ 
         N36 & 100 & $98.90 \pm 0.30 $ \\ 
         N38 & 100 & $98.00 \pm 0.03 $ \\ 
         \bottomrule
    \end{tabular}
\end{table}
The uncertainties are obtained from run-to-run variations across the data-taking period.
The planes with 300~\mum thick sensors have a smaller efficiency of around 93\% due to the selection requirement on the cluster width (see \cref{tab:selection_requirements}).
Removing this selection requirement restores the single-plane efficiencies of these planes from 93\% to 99\% at the cost of a significantly worse spatial resolution. 
The width of the unbiased residuals increases by around 2~\mum or roughly 50\%. 

The efficiency values can also be used to assess possible losses due to higher particle rates than those employed in typical operation.
The number of particles per unit time at SPS is limited and the efficiency studies were performed using dedicated runs.
The particle rate at the SPS H8 line was increased up to \SI{2e6}{} {particles\xspace per\xspace second}, and the relative loss in efficiency for each plane compared to operation at low rate is found to be lower than 1\%.

\section{Conclusion} 
\label{sec:conclusion} 
A charged particle telescope based on the Timepix4 ASIC has been developed and evaluated.
The detector system has been set up to achieve excellent spatial and temporal resolutions for tracking of charged particles, based on eight detector planes equipped with four 100~\mum and four 300~\mum thick silicon sensors. 
The telescope was tested using a 180~GeV/c mixed hadron beam at CERN SPS H8 beamline.

The spatial resolution is evaluated including charge-calibrated pixel weighting and $\eta$ corrections. 
The spatial resolution for the 300~\mum thick sensors resolutions is determined to be $3.3 \pm 3~ \mum$ at \SI{9}{\degree} and ${14.4\pm0.5~\mum}$ for the 100~\mum thick sensors at \SI{0}{\degree}. 
The pointing resolution with this arrangement at the centre of the telescope is 
\XResTrack and \YResTrack in the $x$ and $y$ directions, respectively.  
The temporal resolution of the telescope is measured using per-pixel timing corrections. 
The 300~\mum thick sensors achieve an average resolution of \ThickTimeRes while the 100~\mum thick sensors
achieve \ThinTimeRes.
The combination of the time measurements of the Timepix4 planes yields a temporal resolution of \TrackTimeRes, 
A time resolution of \TRefRes is achieved by the time reference system, composed of a pair of MCPs readout with a PicoTDC, which provides sufficient resolution for very precise timing DUTs.

The efficiency of the telescope is checked as a function of the particle rate. 
The efficiency loss is found to be less than 1\% up to the highest rate available at the SPS, of around \SI{2e6}{} particles per second. 

\section{Acknowledgements}
\label{sec:acknowledgements}
We thank Thierry Gys for guidance and support during the early stages of operation of the MCP-PMTs.
We would like to express our gratitude to our colleagues in the CERN accelerator departments for the excellent performance of the beam in the SPS North Area. 
We also thank the PicoTDC team for providing extensive support in software and the firmware code for modification, as well as the initial prototype boards used in this publication.
We gratefully acknowledge the support of the CERN Strategic R\&D Programme on Technologies for Future Experiments\footnote{For more information, see \url{https://ep-rnd.web.cern.ch/}.} and the computing resources provided by CERN.
We also gratefully acknowledge the support from the following national agencies: the Netherlands Organisation for Scientific Research (NWO), in particular this publication is part of the project FASTER with file
number OCENW.XL21.XL21.076; The Royal Society and the Science and Technology Facilities Council (U.K., grant nos. ST/V003151/1, ST/S000933/1, and ST/W004305/1);
the European Research Council (grant no. 852642);
the Wolfgang Gentner Programme of the German Federal Ministry of Education and Research (grant no. 13E18CHA);
the German Federal Ministry of Education and Research (BMBF, grant no. 05H21PECL1) within ErUM-FSP T04.
This project has received funding from the European Union’s Horizon 2020 Research and Innovation programme under  GA no 101004761.

\clearpage
\addcontentsline{toc}{section}{References}
\setboolean{inbibliography}{true}
\bibliographystyle{elsarticle-num-names}
\bibliography{main}

\begin{thebibliography}{22}
\expandafter\ifx\csname natexlab\endcsname\relax\def\natexlab#1{#1}\fi
\providecommand{\url}[1]{\texttt{#1}}
\providecommand{\href}[2]{#2}
\providecommand{\path}[1]{#1}
\providecommand{\DOIprefix}{doi:}
\providecommand{\ArXivprefix}{arXiv:}
\providecommand{\URLprefix}{URL: }
\providecommand{\Pubmedprefix}{pmid:}
\providecommand{\doi}[1]{\href{http://dx.doi.org/#1}{\path{#1}}}
\providecommand{\Pubmed}[1]{\href{pmid:#1}{\path{#1}}}
\providecommand{\bibinfo}[2]{#2}
\ifx\xfnm\relax \def\xfnm[#1]{\unskip,\space#1}\fi
\bibitem[{Akiba et~al.(2022)}]{LHCbVELOgroup:2022}
\bibinfo{author}{K.~Akiba}, et~al.,
\newblock \bibinfo{title}{{Considerations for the VELO detector at the LHCb
  Upgrade II}},
\newblock
  \bibinfo{journal}{\href{https://cds.cern.ch/record/2800144}{LHCb-PUB-2022-001}}
   (\bibinfo{year}{2022}). \URLprefix \url{http://cds.cern.ch/record/2800144}.
\bibitem[{Akiba et~al.(2019)}]{Akiba:2019faz}
\bibinfo{author}{K.~Akiba}, et~al.,
\newblock \bibinfo{title}{{LHCb VELO Timepix3 Telescope}},
\newblock \bibinfo{journal}{JINST} \bibinfo{volume}{14} (\bibinfo{year}{2019})
  \bibinfo{pages}{P05026}. \DOIprefix\doi{10.1088/1748-0221/14/05/P05026}.
  \href{http://arxiv.org/abs/1902.09755}{{\tt arXiv:1902.09755}}.
\bibitem[{Heijhoff et~al.(2020)}]{Heijhoff:2020mlk}
\bibinfo{author}{K.~Heijhoff}, et~al.,
\newblock \bibinfo{title}{{Timing performance of the LHCb VELO Timepix3
  Telescope}},
\newblock \bibinfo{journal}{JINST} \bibinfo{volume}{15} (\bibinfo{year}{2020})
  \bibinfo{pages}{P09035}. \DOIprefix\doi{10.1088/1748-0221/15/09/p09035}.
  \href{http://arxiv.org/abs/2008.04801}{{\tt arXiv:2008.04801}}.
\bibitem[{Heijhoff et~al.(2021)}]{Heijhoff:2021rtu}
\bibinfo{author}{K.~Heijhoff}, et~al.,
\newblock \bibinfo{title}{{Timing measurements with a 3D silicon sensor on
  Timepix3 in a $180\gevc$ hadron beam}},
\newblock \bibinfo{journal}{JINST} \bibinfo{volume}{16} (\bibinfo{year}{2021})
  \bibinfo{pages}{P08009}. \DOIprefix\doi{10.1088/1748-0221/16/08/P08009}.
  \href{http://arxiv.org/abs/2105.11800}{{\tt arXiv:2105.11800}}.
\bibitem[{Dall'Occo et~al.(2021)}]{DallOcco:2021tjb}
\bibinfo{author}{E.~Dall'Occo}, et~al.,
\newblock \bibinfo{title}{{Temporal characterisation of silicon sensors on
  Timepix3 ASICs}},
\newblock \bibinfo{journal}{JINST} \bibinfo{volume}{16} (\bibinfo{year}{2021})
  \bibinfo{pages}{P07035}. \DOIprefix\doi{10.1088/1748-0221/16/07/P07035}.
  \href{http://arxiv.org/abs/2102.06088}{{\tt arXiv:2102.06088}}.
\bibitem[{Llopart et~al.(2022)}]{tpx4_jinst}
\bibinfo{author}{X.~Llopart}, et~al.,
\newblock \bibinfo{title}{Timepix4, a large area pixel detector readout chip
  which can be tiled on 4 sides providing sub-200 ps timestamp binning},
\newblock \bibinfo{journal}{JINST} \bibinfo{volume}{17} (\bibinfo{year}{2022})
  \bibinfo{pages}{C01044}. \DOIprefix\doi{10.1088/1748-0221/17/01/c01044}.
\bibitem[{Heijhoff et~al.(2022)}]{Heijhoff_2022}
\bibinfo{author}{K.~Heijhoff}, et~al.,
\newblock \bibinfo{title}{{Timing performance of the Timepix4 front-end}},
\newblock \bibinfo{journal}{JINST} \bibinfo{volume}{17} (\bibinfo{year}{2022})
  \bibinfo{pages}{P07006}. \DOIprefix\doi{10.1088/1748-0221/17/07/p07006}.
  \href{http://arxiv.org/abs/2203.15912}{{\tt arXiv:2203.15912}}.
\bibitem[{{CERN Engineering Department}(2019)}]{sps-h8}
\bibinfo{author}{{CERN Engineering Department}},
\newblock \bibinfo{title}{{The H8 Secondary Beam Line of EHN1/SPS}}
  (\bibinfo{year}{{2019}}). \URLprefix
  \url{https://sba.web.cern.ch/sba/BeamsAndAreas/H8/H8_presentation.html}.
\bibitem[{Akiba et~al.(2023)}]{Akiba:2022fnt}
\bibinfo{author}{K.~Akiba}, et~al.,
\newblock \bibinfo{title}{{Reconstruction of charged tracks with Timepix4
  ASICs}},
\newblock \bibinfo{journal}{JINST} \bibinfo{volume}{18} (\bibinfo{year}{2023})
  \bibinfo{pages}{P02011}. \DOIprefix\doi{10.1088/1748-0221/18/02/P02011}.
  \href{http://arxiv.org/abs/2210.01442}{{\tt arXiv:2210.01442}}.
\bibitem[{Ballabriga et~al.(2020)Ballabriga, Campbell, and
  Llopart}]{Ballabriga2020}
\bibinfo{author}{R.~Ballabriga}, \bibinfo{author}{M.~Campbell},
  \bibinfo{author}{X.~Llopart},
\newblock \bibinfo{title}{{An introduction to the Medipix family ASICs}},
\newblock \bibinfo{journal}{Radiat. Meas.} \bibinfo{volume}{136}
  (\bibinfo{year}{2020}) \bibinfo{pages}{106271}.
  \DOIprefix\doi{10.1016/j.radmeas.2020.106271}.
\bibitem[{Ballabriga et~al.(2023)}]{Ballabriga2022}
\bibinfo{author}{R.~Ballabriga}, et~al.,
\newblock \bibinfo{title}{{The Timepix4 analog front-end design: Lessons learnt
  on fundamental limits to noise and time resolution in highly segmented hybrid
  pixel detectors}},
\newblock \bibinfo{journal}{Nucl. Instrum. Meth. A} \bibinfo{volume}{1045}
  (\bibinfo{year}{2023}) \bibinfo{pages}{167489}.
  \DOIprefix\doi{10.1016/j.nima.2022.167489}.
\bibitem[{Haimberger(2023)}]{Haimberger:2023fkr}
\bibinfo{author}{J.~Haimberger},
\newblock \bibinfo{title}{{Fast and radiation hard silicon hybrid detectors for
  the LHC-b Upgrade II}},
\newblock \bibinfo{journal}{Vienna, Tech. U.}  (\bibinfo{year}{2023}).
  \DOIprefix\doi{10.34726/hss.2023.81444}.
\bibitem[{Altruda et~al.(2023)Altruda, Christiansen, Horstmann, Perktold,
  Porret, and Prinzie}]{picotdc}
\bibinfo{author}{S.~Altruda}, \bibinfo{author}{J.~Christiansen},
  \bibinfo{author}{M.~Horstmann}, \bibinfo{author}{L.~Perktold},
  \bibinfo{author}{D.~Porret}, \bibinfo{author}{J.~Prinzie},
\newblock \bibinfo{title}{Picotdc: a flexible 64 channel tdc with picosecond
  resolution},
\newblock \bibinfo{journal}{Journal of Instrumentation} \bibinfo{volume}{18}
  (\bibinfo{year}{2023}) \bibinfo{pages}{P07012}. \URLprefix
  \url{https://dx.doi.org/10.1088/1748-0221/18/07/P07012}.
  \DOIprefix\doi{10.1088/1748-0221/18/07/P07012}.
\bibitem[{Peters et~al.(2015)Peters, Sindrilaru, and Adde}]{Peters:2015aba}
\bibinfo{author}{A.~J. Peters}, \bibinfo{author}{E.~A. Sindrilaru},
  \bibinfo{author}{G.~Adde},
\newblock \bibinfo{title}{{EOS as the present and future solution for data
  storage at CERN}},
\newblock \bibinfo{journal}{J. Phys. Conf. Ser.} \bibinfo{volume}{664}
  (\bibinfo{year}{2015}) \bibinfo{pages}{042042}.
  \DOIprefix\doi{10.1088/1742-6596/664/4/042042}.
\bibitem[{Gaspar et~al.(2001)Gaspar, D\"onszelmann, and
  Charpentier}]{Gaspar:2001fbw}
\bibinfo{author}{C.~Gaspar}, \bibinfo{author}{M.~D\"onszelmann},
  \bibinfo{author}{P.~Charpentier},
\newblock \bibinfo{title}{{DIM, a portable, light weight package for
  information publishing, data transfer and inter-process communication}},
\newblock \bibinfo{journal}{Comput. Phys. Commun.} \bibinfo{volume}{140}
  (\bibinfo{year}{2001}) \bibinfo{pages}{102--109}.
  \DOIprefix\doi{10.1016/S0010-4655(01)00260-0}.
\bibitem[{Krummenacher(1991)}]{Krummenacher:1991qhr}
\bibinfo{author}{F.~Krummenacher},
\newblock \bibinfo{title}{{Pixel detectors with local intelligence: an IC
  designer point of view}},
\newblock \bibinfo{journal}{Nucl. Instrum. Meth. A} \bibinfo{volume}{305}
  (\bibinfo{year}{1991}) \bibinfo{pages}{527--532}.
  \DOIprefix\doi{10.1016/0168-9002(91)90152-G}.
\bibitem[{Jakubek(2011)}]{Jakubek:2011dsm}
\bibinfo{author}{J.~Jakubek},
\newblock \bibinfo{title}{{Precise energy calibration of pixel detector working
  in time-over-threshold mode}},
\newblock \bibinfo{journal}{Nucl. Instrum. Meth. A} \bibinfo{volume}{633}
  (\bibinfo{year}{2011}) \bibinfo{pages}{S262--S266}.
  \DOIprefix\doi{10.1016/j.nima.2010.06.183}.
\bibitem[{Workman et~al.(2022)}]{Workman:2022ynf}
\bibinfo{author}{R.~L. Workman}, et~al. (\bibinfo{collaboration}{Particle Data
  Group}),
\newblock \bibinfo{title}{{Review of Particle Physics}},
\newblock \bibinfo{journal}{PTEP} \bibinfo{volume}{2022} (\bibinfo{year}{2022})
  \bibinfo{pages}{083C01}. \DOIprefix\doi{10.1093/ptep/ptac097}.
\bibitem[{Delogu et~al.(2024)}]{Delogu:2024}
\bibinfo{author}{P.~Delogu}, et~al.,
\newblock \bibinfo{title}{{Validation of Timepix4 energy calibration procedures
  with synchrotron X-ray beams}},
\newblock \bibinfo{journal}{Nucl. Instrum. Meth. A} \bibinfo{volume}{1068}
  (\bibinfo{year}{2024}) \bibinfo{pages}{169716}.
  \DOIprefix\doi{10.1016/j.nima.2024.169716}.
\bibitem[{Akiba et~al.(2012)}]{Akiba:2011vn}
\bibinfo{author}{K.~Akiba}, et~al.,
\newblock \bibinfo{title}{{Charged Particle Tracking with the Timepix ASIC}},
\newblock \bibinfo{journal}{Nucl. Instrum. Meth. A} \bibinfo{volume}{661}
  (\bibinfo{year}{2012}) \bibinfo{pages}{31--49}.
  \DOIprefix\doi{10.1016/j.nima.2011.09.021}.
  \href{http://arxiv.org/abs/1103.2739}{{\tt arXiv:1103.2739}}.
\bibitem[{Blobel(2006)}]{Blobel:2006yh}
\bibinfo{author}{V.~Blobel},
\newblock \bibinfo{title}{{Software alignment for tracking detectors}},
\newblock \bibinfo{journal}{Nucl. Instrum. Meth. A} \bibinfo{volume}{566}
  (\bibinfo{year}{2006}) \bibinfo{pages}{5--13}.
  \DOIprefix\doi{10.1016/j.nima.2006.05.157}.
\bibitem[{Akiba et~al.(2013)}]{Akiba:201347}
\bibinfo{author}{K.~Akiba}, et~al.,
\newblock \bibinfo{title}{{The Timepix Telescope for high performance particle
  tracking}},
\newblock \bibinfo{journal}{Nucl. Instrum. Meth. A} \bibinfo{volume}{723}
  (\bibinfo{year}{2013}) \bibinfo{pages}{47--54}.
  \DOIprefix\doi{https://doi.org/10.1016/j.nima.2013.04.060}.

\end{thebibliography}

\end{document}